\documentclass{osa-article}

\journal{oe}
\usepackage{color}

\newcommand{\bl}[1]{\textcolor{blue}{ #1}}
\usepackage{tikz}
\newcommand\encircle[1]{%
  \tikz[baseline=(X.base)] 
    \node (X) [draw, shape=circle, inner sep=-.5] {\strut #1};}
\articletype{Research Article}

\begin{document}

\title{Single-shot Stokes polarimetry enabled by a digital micromirror device}

\author{Bo-Zhao\authormark{1}, Xiao-Bo Hu\authormark{1}, Valeria Rodr\'iguez-Fajardo\authormark{2}, Zhi-Han Zhu\authormark{1}, Wei Gao\authormark{1}, Andrew Forbes\authormark{2} and Carmelo Rosales-Guzm\'an\authormark{1*}}

\address{\authormark{1}Wang Da-Heng Collaborative Innovation Center for Quantum manipulation \& Control, Harbin University of Science and Technology, Harbin 150080, China}
\address{\authormark{2}School of Physics, University of the Witwatersrand, Private Bag 3, Johannesburg 2050, South Africa}
\email{\authormark{*}carmelorosalesg@hrbust.edu.cn} 


\begin{abstract}
Stokes polarimetry (SP) is a powerful technique that enables spatial reconstruction of the state of polarization (SoP) of a light beam using only intensity measurements. A given SoP is reconstructed from a set of four Stokes parameters, which are computed through four intensity measurements. Since all intensities must be performed on the same beam, it is common to record each intensity individually, one after the other, limiting its performance to light beams with static SoP. Here, we put forward a novel technique to extend SP to a broader set of light beams with dynamic SoP. This technique relies on the superposition principle, which enables the splitting of the input beam into identical copies, allowing the simultaneous measurement of all intensities. For this, the input beam is passed through a multiplexed digital hologram displayed on a polarization-insensitive Digital Micromirror Device (DMD) that grants independent and rapid (20 kHz) manipulation of each beam. We are able to reliably reconstruct the SoP with high fidelity and at speeds of up to 27 Hz, paving the way for real-time polarimetry of structured light.
\end{abstract}

\section{Introduction}
Polarization is a remarkable feature of light that evinces its oscillatory wave nature. The study of polarization can be traced back to the 1600s, but it was only after the seminal work by Young, in 1803, that it was linked to the transverse vibrations of the electric field. Many of the greatest minds of the 19th century, including Malus, Brewster, Arago and Fresnel, contributed enormously to the understanding of polarization, but it was Stokes who established the basis for its modern description \cite{Stokes1852}. The significance of his contribution lies in the introduction of four quantities determined through intensity measurements, known as Stokes parameters, to describe any state of polarization. In essence, the unknown field is projected into a set of three unbiased polarization basis. The intensity reaching the photodetector contains information about the percentage of light in each basis, which is captured by the Stokes parameters, from which the SoP can be reconstructed. Incidentally, the Stokes parameters have been also used to measure the wavefront of propagating structured light beams \cite{Dudley2014}. Traditionally, the required intensities are measured one by one, which represents a major drawback in SP since many applications would benefit from a real-time reconstruction of the SoP. For example, present approaches do not allow the real-time polarization tracking of dynamically changing light fields. Previous attempts to ameliorate this have considered amplitude division, which faces the disadvantages of unbalanced distribution of energy, as well as a mismatch in beam size due to unequal propagation distances \cite{Azzam1982,Fridman2010}.  

Here we put forward an all-digital technique for the real-time reconstruction of polarization. For this, we employ a wavefront splitting approach thus avoiding the issues associated to amplitude division. We take full advantage of digital holography through a computer-controlled Digital Micromirror Device (DMD), of great relevance in the generation of structured light beams \cite{Mirhosseini2013,Hu2018,Ren2015,Chen2015DMD,Mitchell2016,Goorden2014,Lerner2012}. More precisely, our technique consist on splitting the input beam into four independent beams propagating along different paths. For this, a digital hologram containing four multiplexed gratings, with unique spatial frequencies, is displayed on the polarization-insensitive DMD. Remarkably, the DMD enables a rapid ($\sim$20 KHz refresh rate) and independent reconfiguration of each beam, its position or intensity, by simply changing the hologram on the DMD. As a proof-of-principle, we applied our technique to Vector Beams (VB), non-homogeneous SoP that play a key in a myriad of applications \cite{rubinsztein2016roadmap,Hu2019,Ndagano2017,Ndagano2018,Bhebhe2018a,Bhebhe2018b,RosalesReview2018,Zhan2009}. To demonstrate the accuracy of our technique, we compared our experimentally reconstructed SoP with numerical simulations.

\begin{figure}[t]
    \centering
    \includegraphics[width=\textwidth]{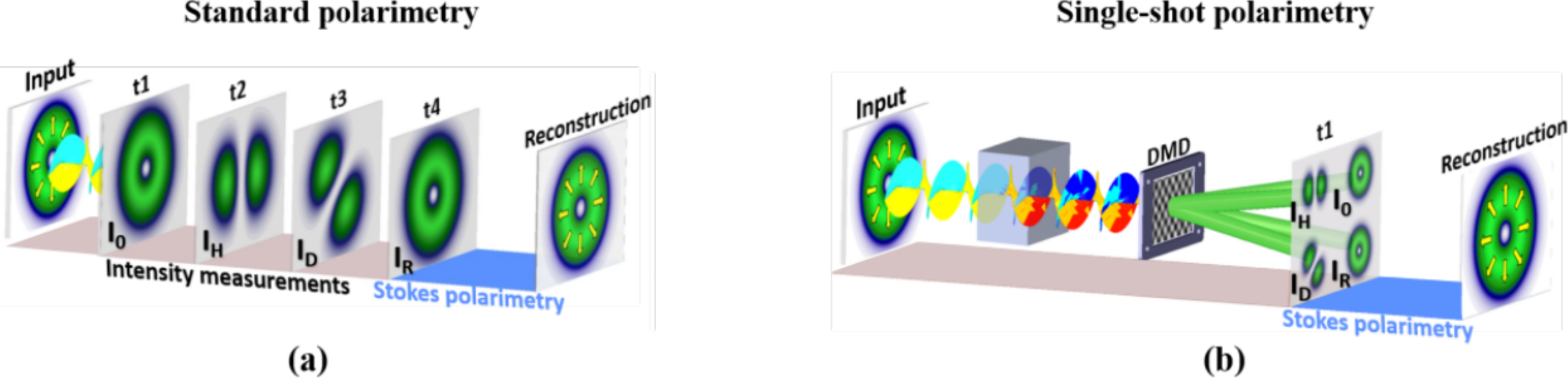}
    \caption{(a) Standard polarymetry: the intensities required to reconstruct a SoP are recorded one by one at times t$_1$, t$_2$, t$_3$ and t$_4$. (b) In our technique, a digital hologram displayed on a Digital Micro Device (DMD), splits the input beams into four identical copies for a simultaneous measurement of the four intensities, at time t$_1$.}
    \label{concept}
\end{figure}
\section{Concept}
In VBs the spatial and polarization degrees of freedom are coupled in a non-separable way. These are commonly expressed as  \cite{Alonso2010,Galvez2012,galvez2014generation},
\begin{equation}
\bf{E}(\rho,\phi)=\cos(\theta)\exp[i\ell\phi]\hat{e_R}+\sin(\theta)\exp[-i\ell\phi]\exp[i\alpha]\hat{e_L},
\label{VectModes}
\end{equation}
where, $(\rho, \phi)$ represent the cylindrical coordinates, the exponential term $\exp[i\ell\phi]$ is associated to an azimuthal phase distribution and $\ell$ $\in \mathbb{Z}$ is known as the topological charge. The amplitude parameter $\theta \in [0, \pi/2]$ allows to change the field $\bf{E}(\rho, \phi)$, from purely scalar ($\theta=0$ and $\theta=\pi/2$) to vector ($\theta=\pi/4$). Finally, the exponential term $\exp[i\alpha]$ is an inter-modal phase. The SoP of such light beams can be reconstructed through the Stokes parameters, which can be computed through a minimum of four intensities measurements as \cite{Stokes1852},
\begin{align}
\centering
     S_{0}=I_{0},\hspace{5mm} S_{1}=2I_{H}-S_{0},\hspace{5mm} S_{2}=2I_{D}-S_{0},\hspace{5mm} \textrm{and}\hspace{5mm} S_{3}=2I_{R}-S_{0},
    \label{Stokes}
\end{align}
where $I_0$ is the total intensity of the beam and $I_H$, $I_D$ and $I_R$ represent the intensity of the horizontal, diagonal and right-handed polarization components, respectively. Figure \ref{concept} (a) illustrates the traditional way in which the SoP of an input beam is reconstructed. Here, the required intensities are recorded at different times (t$_1$, t$_2$, t$_3$ and t$_4$), limiting its applicability to invariant SoP. In contrast, Fig. \ref{concept} (b) illustrates our technique, in which the light beam is projected onto a digital hologram displayed on a polarization-insensitive DMD. The hologram consist of four multiplexed diffraction gratings with unique spatial frequencies that enable the splitting of the input beam into four identical copies. Each of this beams is then passed through the necessary optical filters to measure the required intensities. In this way, all intensities can be recorded in a single image, enabling the reconstruction of the SoP in a single shot, thus allowing real-time tracking of the SoP at speeds limited only by the CCD camera. This technique will be of great relevance in the case when the SoP of the beam changes over time, for example, by passing through a polarization-sensitive system, as schematically illustrated in Fig.  \ref{concept} (b).

\section{Setup}
\begin{figure}[b]
    \centering
    \includegraphics[width=\textwidth]{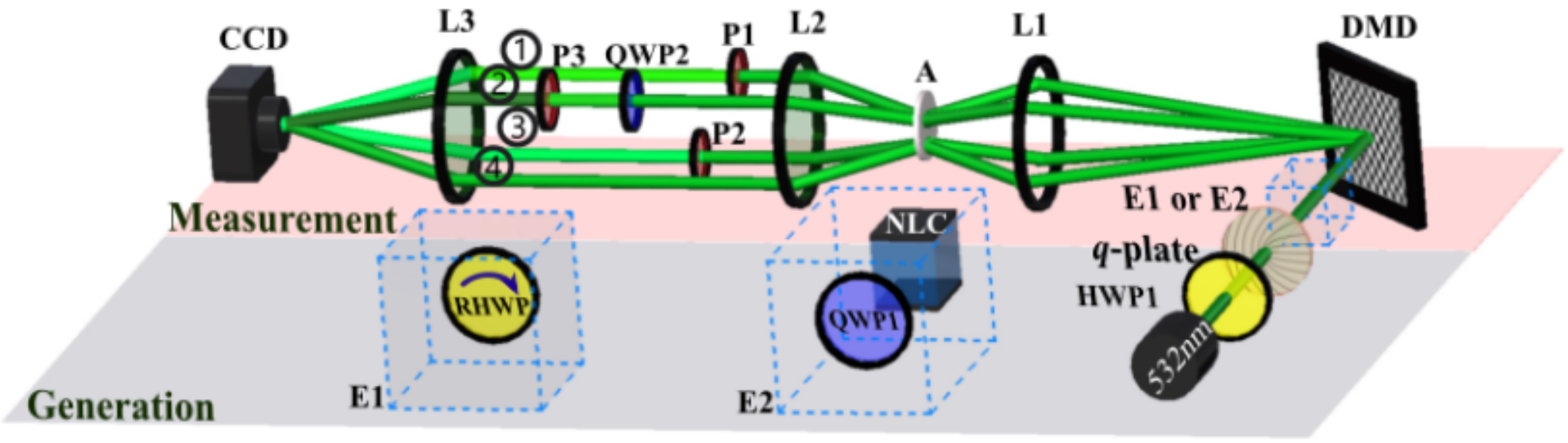}
    \caption{ {\bf Generation} A vector beams is generated from a CW Gaussian beam ($\lambda=532$ nm) using a q-plqte (q=1/2) and a Half Wave-Plate (HWP1). Afterwards, a Digital Micromirror Device (DMD) splits the beam into four identical copies which are propagated along parallel paths using lenses L1 and L2. The linear polarizers P1 and P2 filter $I_H$ and $I_D$, respectively, Whereas P3 in combination with a quarter wave-plate (QWP2) filters $I_R$. Finally, lens L3 focuses the four beams into a CCD to measure all the intensities with a single CCD camera. The systems E1 (a Rotating Half Wave-Plate: RHWP) and E2  (a Quarter Wave-Plate in combination with a non-linear crystal: NLC) enclosed in the dashed boxes were used were inserted in the path of the beam to vary the SoP. }
    \label{setup}
\end{figure}

Our experimental setup is depicted in Fig.\ref{setup}, where a linearly polarized Gaussian beam ($\lambda=532nm$) is converted into a cylindrical vector beam via a q-plate (q=1/2) in combination with a Half Wave-Plate (HWP1) \cite{Marrucci2006}. To reconstruct the SoP of a beam, this is first split into four identical copies using a digital hologram displayed on a DMD (DLP Light Crafter 6500 from Texas Instruments). The generated beams are spatially filtered, to remove higher diffraction orders, and collimated to propagate along parallel paths using lenses L1 and L2 ($f_{1,2}=200$ mm) and the spatial filter (A). Each beam is then used to measure the required intensities $I_0$, $I_H$, $I_D$ and $I_R$ as detailed next. Path \encircle{1} is used to obtain $I_H$ by means of a linear polarizer at $\theta=0^0$ (P1). Path \encircle{3} to measure $I_D$ using a second linear polarizer at $\theta=45^0$ (P2). Path \encircle{2} is used to obtain $I_R$ by means of a QWP (QWP2) at $\beta=45^0$ and a linear polarizer at $\theta=90^0$. Finally, the beam on path \encircle{4} is transmitted unmodified to obtain $I_0$. A third lens (L3, $f_3=200 mm$) is added to focus the beams into a CCD camera (BC106N-VIS from Thorlabs) where the four intensities are measured simultaneously. Importantly, the DMD enables rapid adjustment of each beam's position and intensity, a feature that makes our system very robust compared to others. Once the system is aligned, everything can be programmed to reconstruct the polarization distribution at the click of a button. Additionally, the systems represented as E1 (Rotating Half Wave-Plate) and E2 (Quarter Wave-Plate in combination with a non-linear crystal) were inserted in the path of the beam to change its SoP while we reconstructed it using our system.   

\begin{figure}[tb]
    \centering
    \includegraphics[width=.7\textwidth]{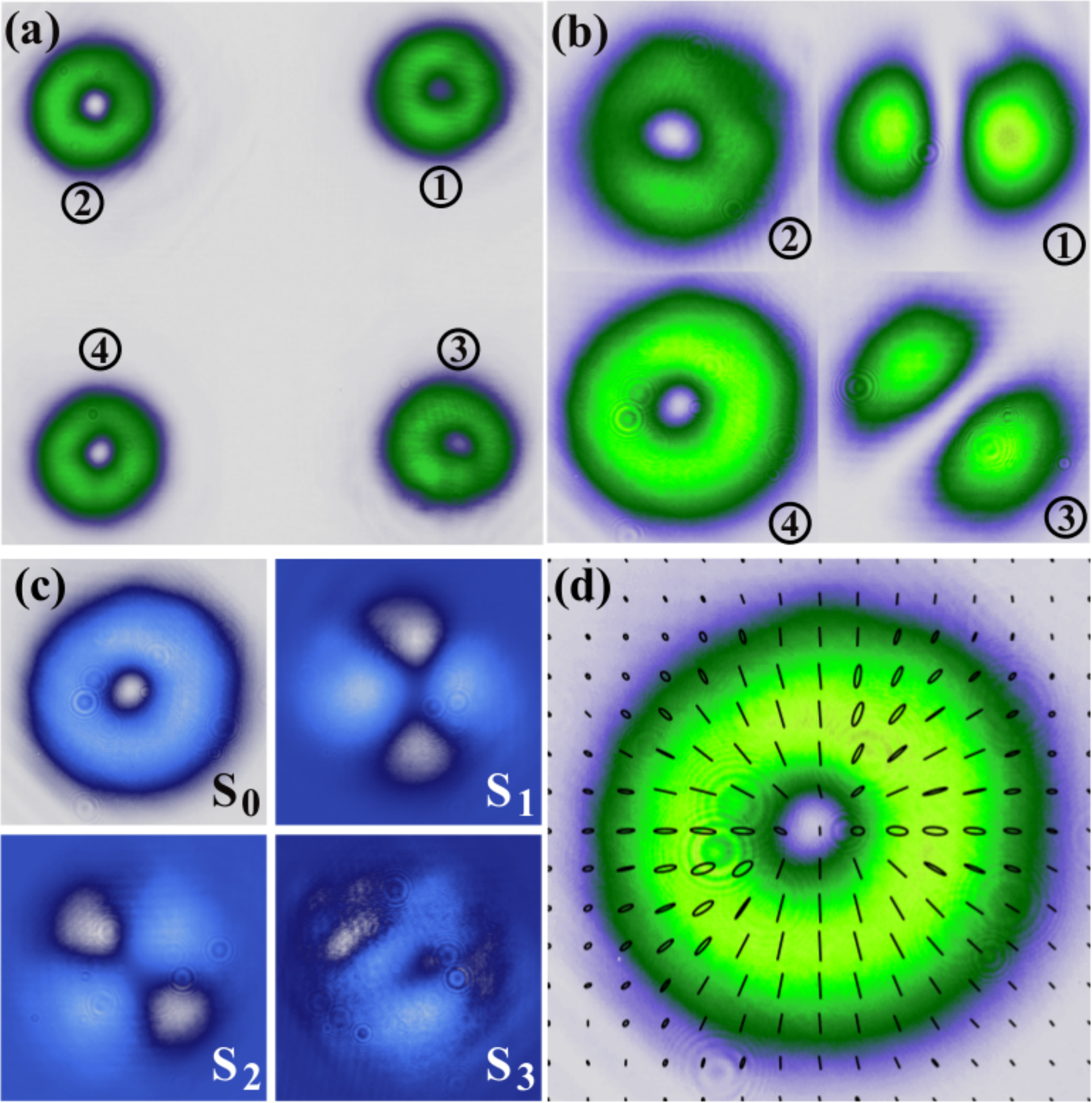}
    \caption{(a) Calibration image to find the centers of the beams. (b) Example image of the intensities $I_0$, $I_D$, $I_H$ and $I_R$. (c) Computed Stokes parameters. (d) Reconstructed polarization distribution.}
    \label{Calibration}
\end{figure}
 To accurately reconstruct the vector beam's polarization using the experimental setup described above, we first recorded a calibration image [see Fig. \ref{Calibration} (a)] that allows to find the center and enclosing area of beams \encircle{1}, \encircle{2}, \encircle{3} and \encircle{4} [see Fig. \ref{Calibration} (b)]. These beams are then used to compute the Stokes parameters, as shown in Fig. \ref{Calibration} (c), from which, the SOP can be accurately reconstructed, as illustrated in Fig. \ref{Calibration} (d), where this is overlapped with the intensity distribution.

\begin{figure}[t]
    \centering
    \includegraphics[width=.75\textwidth]{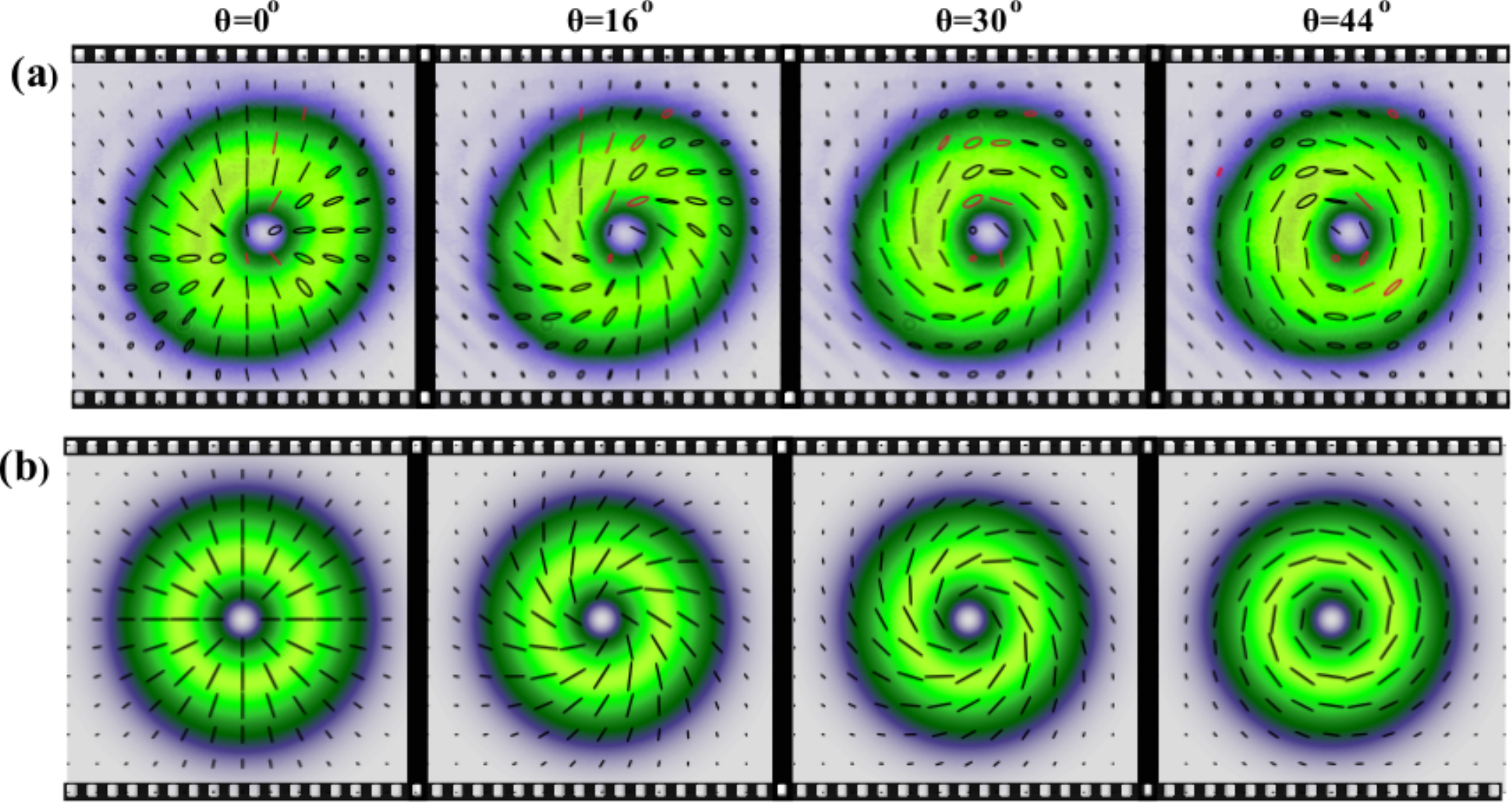}
    \caption{Extracted frames of a real-time polarization reconstruction with a rotating wave plate. (a) Experiment \bl{Visualization 1} and (b) simulation (\bl{Visualization 2})}
    \label{WaveplateRT}
\end{figure}
\begin{figure}[b]
    \centering
    \includegraphics[width=\textwidth]{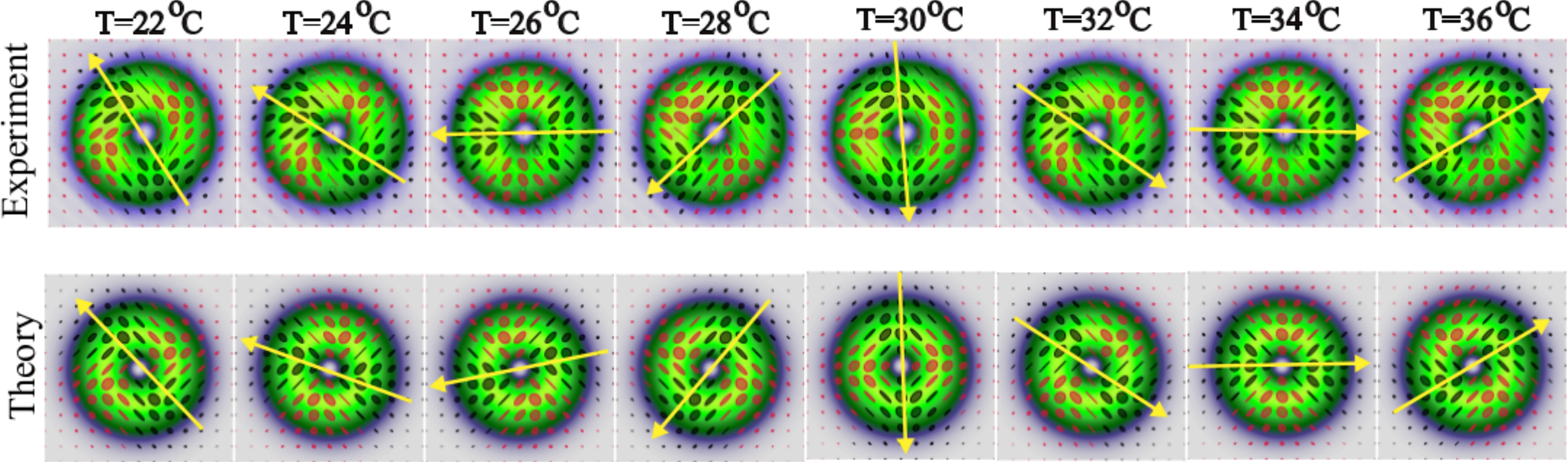}
    \caption{Experimental (top raw) and simulated (botom row) reconstruction of the SoP as the temperature of a non-linear crystal changes from $21 ^\circ C$ to $32^\circ C$. The arrows indicate the rotation of polarization.}
    \label{KTP}
\end{figure}
\section{Results}
As a proof-of-principle, we performed two experiments using the optical systems depicted in the dashed boxes {\bf E1} and {\bf E2} shown in Fig. \ref{setup}. Each system was inserted in the path of the vector beam to change dynamically its SoP while we reconstructed in real time the SoP of the emerging beam. In the first experiment we inserted system {\bf E1} in the path of a radially polarized vector beam, which introduces an angle-dependant phase delay between both orthogonal polarization components. In this way, as we rotate the RHWP at a constant rate, the SoP of the VBs evolves from radial (at $0^\circ$ ) to azimuthal (at $45^\circ$) and back to radial at ($90^\circ$), acquiring intermediate states at intermediate angles. Figures \ref{WaveplateRT} (a) and (b) exhibit six frames of a video showing the experimentally reconstructed (see \bl{Visualization 1}) and numerically simulated (see \bl{Visualization 2}) SoP, respectively, of the VBs as we rotate the RHWP. 

The second experiment was performed with the idea of showing a potential application of our technique, this in the context of optical metrology, for which we used a temperature-controlled birefringent non-linear crystal (NLC), Potassium titanyl phosphate (KTP). The KTP produces a temperature-dependant phase delay between the horizontal ($\hat{H}$) and vertical ($\hat{V}$) polarization components \cite{Zhou2014}. Hence, prior to passing the VBs through the KTP crystal, we first converted the vector beam from the circular to the linear polarization basis using a Quarter Wave-Plate (QWP1) at $45^\circ$ (see inset {\bf E2} in Fig. \ref{setup}). In this way, an increase in the crystal's temperature results in a rotation of the polarization distribution. In the experiment, we increased the temperature from $21 ^\circ C$ to $36 ^\circ C$ while reconstructed the output polarization every $2^ \circ C$. Figure \ref{KTP} the experimentally reconstructed SoP (top row) compared to a numerical simulation of the same (top row) where $\hat{H}$ and $\hat{V}$ polarizations are represented by black and red ellipses, respectively. Here, for the sake of clarity we indicate with a long arrow the center of the $\hat{H}$ polarization, as can be seen, as the temperature increases, the polarization distribution rotates in a counterclockwise direction. Hence, by simply measuring the angle of rotation of the output polarization we can measure the temperature of the crystal.

\section{Conclusions}
Here we have proposed a technique to measure in real-time the SoP of a light beam that takes full advantage of DMD technology. Even though DMDs have been around for almost four decades, it is only in recent times that they started raising attention into the generation of structured light fields. Nonetheless, their capabilities haven't been fully exploited yet, as is the case of their polarization independence. Here we propose to use DMDs as a digital reconfigurable grating to split a input beam into four beams propagating along different paths to perform SP on the input beam to reconstruct its spatial polarization distribution in a single shot. We demonstrated the reconstruction of the SoP in vector beams with high fidelity at speeds limited only by the CCD camera (about 20 Hz). Our technique is not intended for static SoP, even though it can be applied to these, but rather for dynamic SoP. That is the case of VBs in the context of optical communications, either in optical fibers or free-space, where due to external perturbations its polarization distribution changes over time. Hence, a system capable to characterize in real-time the evolution of the SoP is highly desirable, since this will allow to perform real-time correction of the transmitted beam. Another example is in optical metrology, where we can assign a one-to-one correspondence of a given property of a system, for example temperature, to a particular vector state. In this way, we can perform remote sensing of the properties of a system using VBs, which could be advantageous in certain scenarios. Since our technique provides with a way to reconstruct the SoP of a given optical field in real time, it paves the way to novel applications in fields such as optical metrology, as a polarization-based sensor, or in free-space optical communications with vector beams. In addition, our technique can be implemented in a very compact and cost effective way.

\section*{Funding}
 National Natural Science Foundation of China (NSFC) (11574065)

\section*{Disclosures}
The authors declare that there are no conflicts of interest related to this article.

\end{document}